\documentclass[apj]{emulateapj}

\usepackage{graphicx}
\usepackage{natbib}

\newcommand{\ls}{\mathrel{\raise1.16pt\hbox{$<$}\kern-7.0pt 
\lower3.06pt\hbox{{$\scriptstyle \sim$}}}}         
\newcommand{\gs}{\mathrel{\raise1.16pt\hbox{$>$}\kern-7.0pt 
\lower3.06pt\hbox{{$\scriptstyle \sim$}}}}         

\long\def\comment#1{}

\def\fun#1#2{\lower3.6pt\vbox{\baselineskip0pt\lineskip.9pt
  \ialign{$\mathsurround=0pt#1\hfil##\hfil$\crcr#2\crcr\sim\crcr}}}
\def\lap{\mathrel{\mathpalette\fun <}}
\def\gap{\mathrel{\mathpalette\fun >}}

\begin{document}

\title{Improving Photometric Redshifts using GALEX Observations for the SDSS Stripe 82
and the Next Generation of SZ Cluster Surveys}

\author{Michael D. Niemack\altaffilmark{1}, Raul Jimenez\altaffilmark{2},
Licia Verde\altaffilmark{2}, Felipe Menanteau\altaffilmark{3}, Ben
Panter\altaffilmark{4}, David Spergel\altaffilmark{5}}

 \altaffiltext{1}{Department of Physics, Jadwin Hall, Princeton
University, Princeton, NJ 08544, USA; mniemack@princeton.edu}
 \altaffiltext{2}{Institute of Space Sciences (CSIC-IEEC)/ICREA,
Campus UAB, Barcelona 08193, Spain; raul@ieec.uab.es}
 \altaffiltext{3}{Department of Physics and Astronomy, Rutgers University, NJ, USA}
 \altaffiltext{4}{SUPA, Institute
for Astronomy, University of Edinburgh, Royal Observatory, Edinburgh
EH9-3HJ, UK}
 \altaffiltext{5}{Department of Astrophysical Sciences,
Peyton Hall, Princeton University, Princeton, NJ 08544, USA}

\begin{abstract}
Four large-area Sunyaev-Zeldovich (SZ) experiments -- APEX-SZ, SPT,
ACT, and Planck -- promise to detect clusters of galaxies through
the distortion of Cosmic Microwave Background photons by hot ($>
10^6$ K) cluster gas (the SZ effect) over thousands of square
degrees.  A large observational follow-up effort  to obtain
redshifts for these SZ-detected clusters is under way. Given the
large area covered by these surveys, most of the redshifts will be
obtained via the photometric  redshift (photo-$z$) technique. Here
we demonstrate, in an application using $\sim$3000 SDSS stripe 82
galaxies with $r<20$, how the addition of GALEX photometry
($F_{UV}$, $N_{UV}$) greatly improves the photometric redshifts of
galaxies obtained with optical $griz$ or $ugriz$ photometry. In the
case where large spectroscopic training sets are available,
empirical neural-network-based techniques (e.g., ANNz) can yield a
photo-$z$ scatter of $\sigma_z = 0.018 (1+z)$. If large
spectroscopic training sets are not available, the addition of GALEX
data makes possible the use simple maximum likelihood techniques,
without resorting to Bayesian priors, and obtains
$\sigma_z=0.04(1+z)$, accuracy that approaches the accuracy obtained
using spectroscopic training of neural networks on $ugriz$
observations. This improvement is especially notable for blue
galaxies. To achieve these results, we have developed a new set of
high resolution spectral templates based on physical information
about the star formation history of galaxies. We envision these
templates to be useful for the next generation of photo-$z$
applications. We make our spectral templates and new photo-$z$
catalogs available to the community at
www.ice.csic.es/personal/jimenez/PHOTOZ.
\end{abstract}


\section{Introduction}
Thousands of square degrees of the sky are currently being observed
at mm wavelengths by three experiments: APEX-SZ\footnote{{
www.apex-telescope.org}}, South Pole Telescope (SPT)\footnote{{
spt.uchicago.edu}} and  Atacama Cosmology Telescope (ACT)\footnote{{
www.physics.princeton.edu/act}}. In addition, the Planck satellite,
to be launched this fall, will observe the whole sky. The promise of
these surveys is to provide  a nearly mass-selected galaxy cluster
sample  via the Sunyaev-Zeldovich (SZ) effect \citep{SZ}. Because of
the lack of sensitivity of the SZ-effect to redshift, clusters or
groups of galaxies detected this way need follow-up observations at
other wavelengths to determine their redshifts. The large area of
sky covered and the large number of expected detections make
spectroscopic follow-up of galaxies in every cluster prohibitive.
Upcoming surveys will rely on redshifts obtained from broad-band
photometry (photometric redshifts or photo-$z$) or custom-designed
narrow-band photometry \citep{alhambra, pau}. As broad-band
photometry provides low resolution spectral information, the
determination of galaxy-redshifts can be affected by relatively
large errors. Photo-$z$ errors can limit the accuracy of
cosmological studies using galaxies or clusters, which highlights
the importance of improving photo-$z$ determinations. For example,
SPT and ACT will attempt to constrain the dark energy equation of
state using SZ selected clusters and photo-$z$ \citep{SPT_white}.
\cite{lima-hu07} calculate that a photo-$z$ bias of 0.003 and
scatter of 0.03 will cause a $\sim$10\% increase in the amplitude of
the equation of state error bars achieved by SPT using this
approach. The above surveys will require photo-$z$ not only for SZ
clusters but also for field galaxies, to carry out ancillary science
such as exploiting the signal of CMB weak lensing (e.g.,
\citealt{melita}) by large scale structure and the kinetic-SZ effect
(e.g., \citealt{carlosksz}): two powerful probes of the growth of
structure, which are useful, for example, in distinguishing between
modified gravity and dark energy as the source of the present
accelerating expansion of the universe.

The use of broad-band photometry to determine redshifts is not new
(see first attempts by \citealt{Baum} and \citealt{Koo}). In its
minimalistic approach it consists of simply finding the best fit
redshift using a series of galaxy templates, which can be either
chosen from stellar population models or empirically \citep{koo99}
as long as the set is exhaustive (i.e. fully describes the galaxy
population). With the arrival of large spectrographs, it became
clear that a refinement of the above technique could be achieved by
using small subsets of spectroscopic redshifts as ``training sets''
for larger photometric samples. One can then use these training sets
as inputs for empirical fits to the magnitudes versus $z$ (e.g.
\citealt{Budavari_photoz_2005}) or for artificial neural network
codes to compute photo-$z$
\citep{vanzella04,collister04,ANNz_SDSS_07}. Another approach is to
use prior information about galaxies, like the fact that faint
galaxies tend to be farther away, as a Bayesian prior for computing
the redshift likelihood from the templates
\citep{bpz,CFHT_photoz06,zebra}. Other recently developed techniques
that go beyond simple photometry fits include using structural
properties of galaxies like their size or surface brightness to
obtain more accurate photo-$z$ (e.g., \citealt{WrayGunn07}).

The above methods have their pros and cons. For example, methods
based on training sets, because of their empirical basis, can only
be reliably extended as far as the spectroscopic redshift limit.
Training sets for surveys such as the dark energy survey (DES) and
the Large Synoptic Survey Telescope (LSST) survey will need of the
order of hundreds of thousands of spectroscopic redshifts
\citep{connollylsst, frieman}.

To use  Bayesian prior-based methods, one needs to construct and
test different priors for different redshift ranges and surveys,
which also requires spectroscopic redshifts to accurately generate
the prior distributions.

Given the need to obtain relatively accurate photo-$z$ for the large
SZ survey areas we have explored an alternative approach.
The goal of this approach is to optimize photo-$z$ accuracy while
minimizing external assumptions (priors) and additional data
acquisition.

Our approach, presented in detail below, consists of obtaining
moderate depth observations with the Galaxy Evolution Explorer
(GALEX) combined with optical $griz$ data. This data combination was
first tried by \cite{Budavari_photoz_2005}, \cite{Way_photoz_2006},
and \cite{Ball_photoz_GALEX_2007}, who used empirical approaches
with spectroscopic training sets for photo-$z$ determination. Adding
the two GALEX broad bands at central wavelengths of
$\sim$$1500$~\AA\ ($F_{UV}$) and $\sim$$2300$~\AA\ ($N_{UV}$) to
optical $griz$ photometry, improves photo-$z$ determinations, while
requiring minimal assumptions about external priors, for the
following reason. The $4000$~\AA\ break, which is the most commonly
used spectral feature for optical photo-$z$ determination, is
greatly reduced for blue galaxies, making it more difficult to use
as a redshift indicator. This problem is particularly acute at $z
\gap 0.5$, where most galaxies are young and have high star
formation rates (e.g., \citealt{heavens04}). The $912$~\AA\
Lyman-limit, on the other hand, is exhibited by all galaxies
(Fig.~\ref{fig:templates}). Filters that sample closer to the
Lyman-limit help to pin down the galaxy type and redshift,
especially for blue galaxies with no substantial $4000$~\AA\ break.
Further, given the strong sensitivity of the $UV$ to star formation,
one can directly obtain a measure of star formation.

In carrying out this work, we found that galaxy templates with well
motivated blue spectra (in particular, blue-wards of 3000~\AA) are
not publicly available. Either this region of the spectrum was
missing (like in the original \citealt{cww_templates80} templates)
or it was modeled roughly with no spectral features beyond the
Lyman-limit.  Motivated by the need to provide reliable empirical
templates in this region of the spectrum and higher spectral
resolution than currently available models, we have developed our
own templates. To do so we have exploited our knowledge of the star
formation history of galaxies over cosmic time \citep{Panter+07} to
help us build physically motivated templates.

We present a test of the performance of this approach on
spectroscopic samples from  the Sloan Digital Sky Survey (SDSS)
stripe 82 region. We find that, in the case where large
spectroscopic training sets are available, empirical
neural-network-based techniques (e.g., ANNz
\citealt{collister04,ANNz_SDSS_07}) give a $\sigma_z = 0.018(1+z)$
for optical photometry combined with GALEX observations. If large
spectroscopic training sets are not available, the addition of GALEX
data make possible the use of simple maximum likelihood techniques,
without resorting to Bayesian priors, and obtains $\sigma_z =
0.04(1+z)$, which approaches the accuracy obtained using
spectroscopic training of neural networks on $ugriz$ observations.
In particular, we show how the large number of catastrophic failures
that occur for $griz$-based and $ugriz$-based maximum likelihood
photometric redshift determinations is nearly eliminated by adding
UV photometry from GALEX data. The improvement is especially notable
for blue galaxies with $g-r < 0.6$, for which photo-$z$ scatter of
$0.03 (1+z)$ is achieved on galaxies with $r<19$ and $z \lap 0.25$.
As noted below and by \cite{CFHT_photoz06}, the absence of the $u$
band significantly degrades the performance of the photo-$z$
estimation. We show that the addition of GALEX $UV$ observations is
preferable to the addition of optical $u$ band observations.

The rest of the paper is organized as follows: in \S 2 we describe
our source sample. In \S 3 we present our method and details of the
implementation.  In \S 4 we discuss our results. Conclusions are
presented in \S 5.

\section{Source sample}

Our GALEX observations comprise a Legacy program awarded in cycle 3,
with the goal of mapping $\sim$100 deg$^2$ with 3~ks exposure time
per pointing in both the $F_{UV}$ and $N_{UV}$ filters. We chose to
map roughly 11 deg$^2$ covering the Blanco Cosmology
Survey\footnote{cosmology.uiuc.edu/BCS} (BCS) 23-hour field at
declination -55$^\circ$ and a larger area of the equatorial stripe
82, which is covered by SDSS. Both areas have $griz$ observations,
and SDSS also has $u$ observations. The BCS field is part of the
common SZ area survey; however, as there is currently no significant
sample of spectroscopic redshifts in the BCS region to validate our
photo-$z$, BCS analysis will not be presented here. The SDSS stripe
82 has been observed by ACT and (of course) offers a sample of SDSS
spectroscopic redshifts to test the photo-$z$ performance. We took
advantage of the fact that the stripe 82 survey area includes a
number of the GALEX Medium Imaging Survey (MIS) fields, which
already had many $>1.5$ ks observations and therefore needed only
partial additional observations to reach our 3 ks target. In total
we will collect $\sim$210 ks of integration time
--- merely 2.4 days of observations.

\begin{figure}
\begin{centering}
\includegraphics[width=0.9\columnwidth]{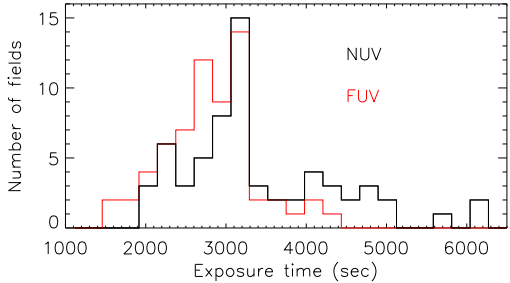}
\caption{Distribution of exposure times on the GALEX fields used in
this analysis for $N_{UV}$ (black) and $F_{UV}$ (red) observations.
Fields were only used with $N_{UV}$ exposure $>2$ ks.}
\label{fig:exposure_times}
\end{centering}
\end{figure}

At the time this analysis was completed, only about half of the
planned observations had been made. The stripe 82 data set used for
this analysis is comprised of 56 GALEX fields ($\sim$55 deg$^2$ of
coverage, although some field edges lie outside the SDSS stripe 82
region) to $N_{UV}$ depths between 2 ks and 6.5 ks
(Fig.~\ref{fig:exposure_times}). These depths allow us to probe
deeper magnitudes and a more complete sample than has been possible
with previous photo-$z$ studies that used GALEX data
\citep{Budavari_photoz_2005,Way_photoz_2006,Ball_photoz_GALEX_2007}.
Of those fields, 41 are publicly available MIS data, and the other
15 are from our guest investigator
proposal.\footnote{\cite{Niemack_thesis} describes the source sample
in more detail.}

\subsection{Magnitudes}

Accurate photometry is critical for obtaining accurate photo-$z$.
Because of the differences in the point spread functions (PSF) of
different instruments and between bands, simple aperture photometry
is not appropriate for this study. The SDSS PSF widths are
approximately 1.5$''$ and vary with sky brightness
\citep{ABAZAJIAN_SDSS_2003}, while GALEX PSF widths vary across the
field between roughly 4$''$ and 7$''$ \citep{Morrisey_GALEX_05}. Our
approach is to use AB magnitude measures that are as close as
possible to the total flux emitted by the galaxy in each band.

As part of the standard GALEX pipeline for each field, SExtractor is
run on both the $F_{UV}$ and $N_{UV}$
images\footnote{galex.stsci.edu/GR2/?page=ddfaq\#2} to extract
multi-pixel sources that are detected above the noise threshold in
background-subtracted images \citep{Sextractor1996}. We use the
$N_{UV}$ and $F_{UV}$ {\it mag\_auto} outputs of SExtractor, which
optimizes elliptical apertures for each source to integrate the
total flux. The $F_{UV}$ bandwidth and transmission are both roughly
a factor of two smaller than the $N_{UV}$
(Figure~\ref{fig:templates}), causing it to have substantially lower
sensitivity. Because of this, far fewer sources are independently
detected in the $F_{UV}$ band.

For the SDSS data we explored the use of both {\it C-model} and {\it
model} magnitude measurements. These magnitudes consist of fitting
models to the profile of the galaxy composed of an exponential disc
and a deVaucouleurs profile. The fits are integrated to three and
seven times the characteristic radius respectively, at which point
the function is truncated to smoothly go to zero within one
additional characteristic radius. For the {\it model} magnitudes all
bands are measured using the best fit model to the $r$-band data,
while for the {\it C-model} magnitudes, the two fits are weighted
based on the quality of the fit and combined to obtain the best
fitting profile for each filter
band.\footnote{www.sdss.org/dr5/algorithms/photometry.html} The {\it
C-model} measurement provides the best estimate of the total
photometric flux for each SDSS band.\footnote{While it may not be
the case for low signal-to-noise cases, in the high signal-to-noise
regime this procedure yields the total photometric flux and is not
affected by systematic errors. For the objects considered the SDSS
signal-to-noise is $> 5$.} While testing the template-fitting
photo-$z$ techniques (\S \ref{sec:SED_photoz_anal}) we found that
{\it model} magnitudes provide a better relative calibration when
comparing only SDSS bands (especially after adding the
\cite{Padmanabhan_Ubercal_2007} ``ubercalibration'' corrections);
however, the {\it C-model} magnitudes provide a better absolute
calibration for comparing with other instruments, such as GALEX.
Both {\it model} and {\it C-model} magnitudes were also tested using
the ANNz analysis described in \S \ref{sec:ANNz_anal}, and no
significant differences were found between the results using the
different magnitudes. The ANNz results presented in \S
\ref{sec:ANNzG_results} were calculated using {\it model}
magnitudes.

Magnitude corrections of -0.04 and +0.02 are applied to the SDSS $u$
and $z$ bands respectively to convert from SDSS magnitudes into AB
magnitudes.\footnote{www.sdss.org/dr6/algorithms/fluxcal.html\#sdss2ab}
All reported magnitudes are in the AB system. In \S
\ref{sec:Results} we assess the performance of our photo-$z$
analysis on those SDSS galaxies with spectroscopic redshifts with
confidence $> 0.9$. SDSS objects are excluded from the catalog using
the ``blended,'' ``nodeblend,'' and ``saturated'' flags. The
majority of the SDSS spectroscopic measurements have $r < 18$,
although, there are also a substantial number of spectroscopic
measurements between $18<r<20$ (which are primarily Luminous Red
Galaxies), so we have limited our current analysis to the $r<20$
magnitude regime (except as discussed in
Fig.~\ref{fig:SDSS_ANN_BPZ_compare} and \S
\ref{sec:Photoz_catalog}).

We emphasize that by using total magnitudes for each band we
minimize the potential problem of missing light because of choosing
an aperture in one band that does not encompass all the light in
other bands. Measuring the total light is important when using the
GALEX bands both because of the different PSF sizes and because most
of the star formation in galaxies takes place at the galaxy
perimeter; thus, a fixed aperture based on a single optical band can
exclude much of the light from recent star formation, which is
measured by the $UV$ bands.

\subsection{Catalog matching}

The GALEX and optical catalogs are merged as follows: we initially
assign optical sources to a GALEX field pointing if they fall within
35.1$'$ of the GALEX field center. This cuts the noisiest region of
the GALEX fields (near the edges), while maintaining complete sky
coverage between neighboring fields (i.e. leaving no gaps). Within
every GALEX field, each optical source is matched to the nearest
GALEX object with a $N_{UV}$ detection within a 4$''$ radius, which
is a relatively conservative matching radius
\citep{Agueros_GALEX_SDSS_2005}. After all sources in the field are
assigned, the combined catalog is searched to test whether any two
optical sources are assigned to the same GALEX object. When there
are overlapping assignments, the closest source to the GALEX
position is selected and the other is removed from the
catalog.\footnote{Removing sources with overlapping assignments was
also explored and had negligible impact on the results in this paper
since so few sources had overlapping assignments.} Sources that do
not have a GALEX detection or overlapping assignments are kept in
the catalog for spectroscopic confirmation tests. We characterize
the distributions of GALEX $F_{UV}$ and $N_{UV}$ magnitudes in each
field using histograms with 0.1 magnitude bins. The magnitude limit
used for other sources in the same field during photo-$z$ analysis
is set to be the highest magnitude where the number of galaxies
exceeds half of the number at the peak magnitude bin
\citep{Niemack_thesis}. Sources with magnitudes higher than this
limit (as well as objects with no $N_{UV}$ detection) are labeled as
non-detections, and this magnitude limit is used for the
non-detections in the photo-$z$ calculation (\S
\ref{sec:photoz_methods}).

In 56 GALEX fields in stripe 82, $\sim$3000 SDSS sources with
spectroscopic redshifts were found that meet the above criteria. Of
these sources, 75\% were found to have $N_{UV}$ detections within
the 4$''$ matching radius, and only two pairs of them were matched
to the same GALEX source.  Only seven of the sources had $r > 20$,
which we treat as the magnitude limit of the spectroscopic analysis.
When generating the photo-$z$ catalogs, we also use SDSS objects
without spectroscopic data. In the same 56 GALEX fields almost
150,000 SDSS sources were found with magnitude $r<21$, and 55\% of
those were successfully matched to GALEX sources. Less than 1\% of
those were excluded because they were matched to the same GALEX
source as another optical source. Both catalogs were also searched
for SDSS sources with multiple GALEX sources within a 4$''$ radius,
and none were found in the spectroscopic catalog, while nine were
found in the photometric sample and were removed from the catalog.

\section{Methods} \label{sec:photoz_methods}

After adding the GALEX bands, we consider two different approaches
for computing the photo-z. First we describe a spectral energy
distribution (SED, or template-based) photo-$z$ calculation
technique and the new SED templates that we have developed. This
approach assumes no prior knowledge of the redshift distribution and
does not require spectroscopic measurements. Our second approach is
to analyze the same GALEX plus optical catalog using ANNz techniques
to train the photo-$z$ calculation with the SDSS spectroscopic
measurements.

To quantify the accuracy of different photo-$z$ analyses,  we define
the redshift error as
\begin{equation}
dz \equiv \frac{(z_{ph} - z_{sp})}{(1+z_{sp})},
\end{equation}
where $z_{ph}$ is the photo-$z$ and $z_{sp}$ is the spectroscopic
$z$. The mean, $z^{bias}$, and standard deviation, $\sigma_z$, of
$dz$ (i.e. the photo-$z$ bias and scatter)  are calculated for all
galaxies with $z_{ph}<1$, which is motivated by the fact that given
the optical and $UV$ depths, we do not expect to detect galaxies
near or above $z=1$. In the SDSS results presented, these $z>1$
failures amount to less than 1\% of the galaxies in the
spectroscopic catalog and $\sim$1\% of the GALEX detected galaxies
with $r<20$ in the photometric catalog. A final cut is made on
objects with $N_{UV} - g > 1$ as this color is typical of QSO's
rather than galaxies. This cut also removes less than 1\% of the
complete SDSS spectroscopic catalog and $\sim$3\% of the GALEX
detected galaxies with $r<20$ in the photometric catalog.

The analysis is done on different combinations of the seven optical
(SDSS) and $UV$ bands. This allows us to study the impact of
including different bands on photo-$z$ accuracy and thus estimate
the importance of different bands for future observations. Our
photo-$z$ are then compared to the recently published results of the
SDSS ANNz photo-$z$ pipeline (henceforth ANNz;
\citealt{ANNz_SDSS_07}), which was developed using a spectroscopic
training and validation set comprised of $\sim$640,000 galaxies
(Fig.~\ref{fig:ML_comparisons}).

We note that when reporting standard deviations to study the
performance of the photo-$z$ we have {\em not} excluded outliers
(with the exception of cutting the small number of galaxies with
$z>1$). Excluding or down-weighting outliers is common practice in
the photo-z literature, motivated by the fact that the photo-$z$
error distribution often is a Gaussian around the peak but has long
tails. As we quote standard deviations with the outliers included,
caution is needed when comparing our numbers with those in the
literature. In particular, for the maximum likelihood analysis
presented in Fig.~\ref{fig:ML_comparisons}, the standard
down-weighting of the outliers would reduce the $ugriz$ photo-$z$
scatter by 20\% and the GALEX + $griz$ photo-$z$ scatter by 15\%,
while having a much smaller effect on the SDSS ANNz performance.

\subsection{New spectral templates for photo-$z$}

For template-based photo-$z$ calculation, we need a basis of
spectral templates that represents galaxies in the redshift range of
interest and for the magnitude range of the catalog. The approach in
the literature so far has been to either use empirical templates
\citep{cww_templates80,Kin_templates96} or use synthetic models with
simple receipes to model the star formation law in galaxies, most
typically using a declining exponential.

\begin{figure}
\begin{centering}
\includegraphics[width=\columnwidth]{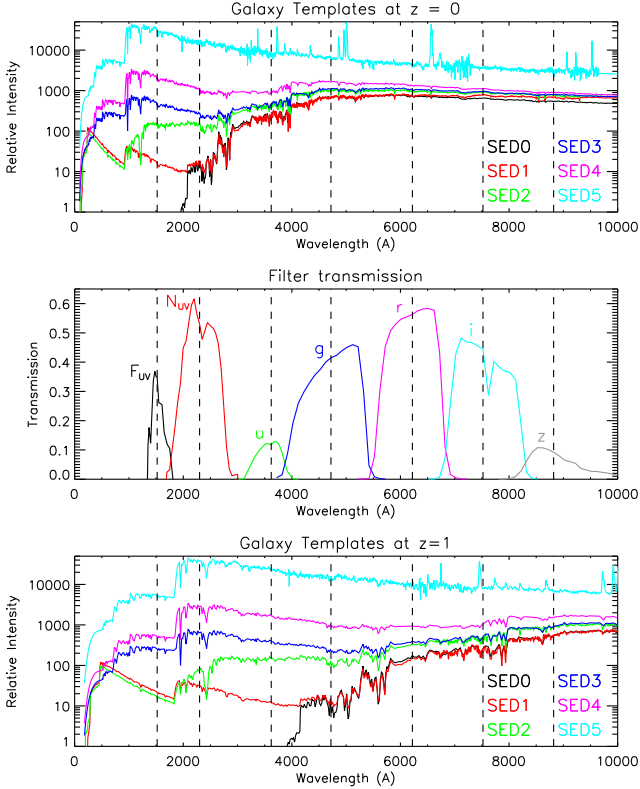}
\caption{The top panel shows the six galaxy templates that are used
to find the maximum likelihood solution for the photometric
redshifts (\S \ref{sec:SED_photoz_anal}). The vertical dashed lines
show the central frequencies of the GALEX and SDSS bandpasses. The
middle panel shows the two GALEX bands ($F_{UV}$, $N_{UV}$) as well
as the five SDSS bands ($u$, $g$, $r$, $i$, $z$). The bottom panel
shows the templates redshifted to $z=1$. As the different galaxy
types are redshifted, a redshift-brightness degeneracy arises in the
optical bands (especially when only considering $griz$ bands) for
the galaxies with blue spectra. The addition of the GALEX bands
breaks this degeneracy by sampling out to the 912~\AA\ Lyman-limit.
Note that by $z=1$ the Lyman-limit has shifted out of the $F_{UV}$
band, but it does not reach the central frequency of the more
sensitive $N_{UV}$ band until $z\approx1.5$.} \label{fig:templates}
\end{centering}
\end{figure}

Recent advances in both observations and stellar modeling have
allowed different groups to determine the complete star formation
history of galaxies \citep{heavens04,fernandes2005,Panter+07} for a
wide range of galaxy stellar masses ($10^7 - 10^{12}$ M$_{\odot}$).
\citet{Panter+07} found that stellar mass is the parameter that most
directly determines the galaxy's star formation history and SED.
Taking advantage of this finding, we use six mass ranges with their
corresponding reconstructed star formation histories, to obtain six
spectral templates. These templates should encompass the entire
galaxy population and are therefore a representative basis of
galaxies in the universe. The spectral templates are built using the
input star formation history with solar metallicity (changing the
template metallicity has little impact on the final photo-$z$
performance) using the \citet{BC03} models. The models have only
absorption lines, so for the star-bursting galaxy template (SED5 in
Fig.~\ref{fig:templates}) we use the emission lines from the
\citet{Kin_templates96} models and add them to this template only.
The new templates -- shown in Fig.~\ref{fig:templates} -- provide a
higher resolution and wider spectral range than other publicly
available templates. Note that we have not adjusted the templates to
obtain the best photometric redshifts, but rather we have used the
physical knowledge of the recovered star formation history of the
universe as our input. We evaluate the performance of these new
templates below.

\subsection{Template fitting photo-$z$ calculation} \label{sec:SED_photoz_anal}

With the addition of the two GALEX bands, our template-based
methodology to obtain photo-$z$ is fairly simple. We use the six
galaxy templates in Fig.~\ref{fig:templates} and perform a maximum
likelihood (ML) analysis, which is simply a chi-square minimization,
to find the best fitting model to the observed photometry. No priors
are used for this analysis, or more accurately, we assume flat
redshift and template priors. We use two codes to compute the
photo-$z$: BPZ\footnote{Code version bpz.1.98b;
acs.pha.jhu.edu/$\sim$txitxo/bpzdoc.html} \citep{bpz} which has the
ability to simultaneously calculate photo-$z$ using both ML analysis
and a Bayesian prior for comparison, and
ZEBRA\footnote{www.exp-astro.phys.ethz.ch/ZEBRA/} \citep{zebra}
which is a recently released independent code that uses similar
techniques to BPZ. Most of our analysis will be done using BPZ
because when including the $UV$ data it performs significantly
better than ZEBRA on our data set in the redshift range $0.25 < z <
0.4$; although, we note that slightly better results can be obtained
by ZEBRA at $z < 0.25$.\footnote{The same differences between BPZ
and ZEBRA were observed with a variety of galaxy templates when
using the GALEX data; however, when the optical data is analyzed
without GALEX data, BPZ and ZEBRA provide nearly identical results.}

The observed magnitudes are matched to the predicted spectral energy
distributions through each bandpass from the templates in
Fig.~\ref{fig:templates}. As suggested by \cite{bpz}, two points of
interpolation are allowed between the different templates in color
space, which allows the best fit template to be a (2:1 or 1:2) mix
of two neighboring templates. The photo-$z$ computation is set to
have a precision of $\delta z = 0.01$. The only limit imposed in the
ML calculation is a sharp prior $z < 1.5$; further, (as described
above) we exclude from the sample the small number of sources with
photo-$z$ $>1$.

\subsection{Artificial neural-network photo-$z$ calculation}
\label{sec:ANNz_anal}

We also consider the empirical photo-$z$ method ANNz developed by
\cite{collister04}. We compare the performance of our template-based
photo-$z$ method to the results of \citet{ANNz_SDSS_07}, who trained
and validated their artificial neural network on 640,000 galaxies
with $ugriz$ SDSS photometry and provide a photo-$z$ catalog for
SDSS galaxies with $r<22$. We use their photo-$z$ determinations for
the galaxies in our sample as a benchmark to compare the performance
of our template-based technique (\S \ref{sec:SDSS_spec} and
Fig.~\ref{fig:ML_comparisons}, \ref{fig:zerr_z},
\ref{fig:zerr_r_mag}, and \ref{fig:zerr_gr}). ANNz in this case
yields a photo-$z$ scatter of $\sigma_z=0.027(1+z)$.

To explore the photo-$z$ potential of GALEX observations in more
detail, we also use the publicly available ANNz
code\footnote{zuserver2.star.ucl.ac.uk/$\sim$lahav/annz.html} with
our combined GALEX and SDSS catalog to obtain more accurate
photo-$z$ (henceforth ANNzG). We use as a training set the SDSS
galaxies in our GALEX fields that also have spectra. Of the
$\sim$$3000$ objects with SDSS spectroscopic redshifts in our
catalog $700$ are used as our training set and $400$ as our
validation set. We then re-run ANNz on the full $\sim$$3000$ objects
to estimate its performance (\S \ref{sec:ANNzG_results}). Two
different network architectures were explored for the ANNzG
analysis: one with five hidden layers with 10 nodes each and three
committee members, and a simpler version with two hidden layers with
10 nodes each and no committee. The more complex architecture did
result in a slight reduction of the photo-$z$ scatter; however, the
relative results of using different data combinations were nearly
identical. We present the results of the simpler network analysis
here. Note that all galaxies in the specified magnitude range are
included in the ANNz and ANNzG analysis, since (unlike the
template-based analysis) there are no galaxies with photo-$z>1$, and
because of the nature of the ANNz calculation, it can also
simultaneously calculate the photo-$z$ for the bluest objects, such
as QSOs.

\section{Results} \label{sec:Results}

\subsection{Photo-$z$ analysis with no priors} \label{sec:SDSS_spec}

\begin{figure*}
\begin{centering}
\includegraphics[width=1.76\columnwidth]{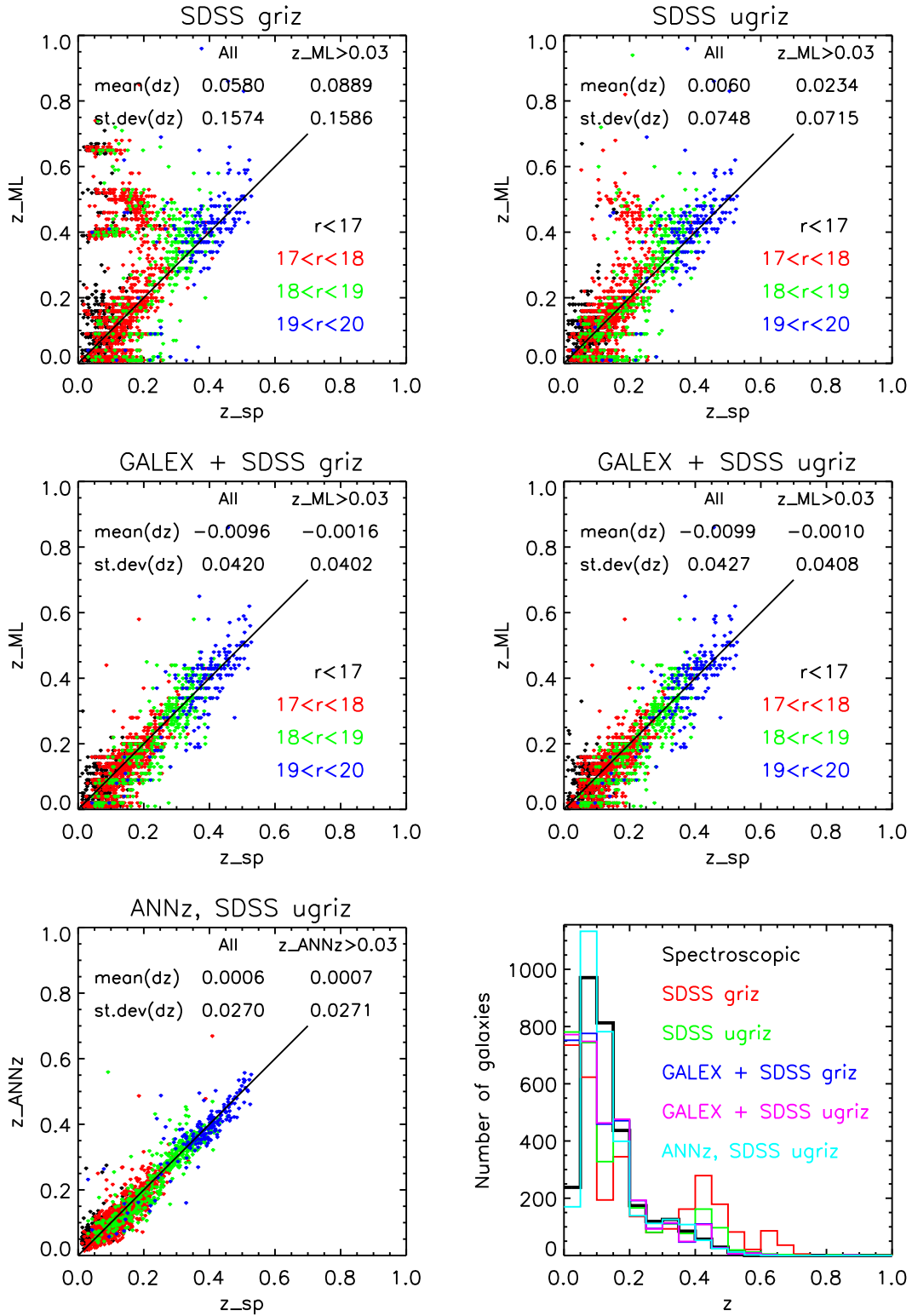}
\caption{Comparison of our maximum likelihood photo-$z$ redshift
estimates ($z$\_ML) and the ANNz estimates ($z$\_ANNz) with SDSS
spectroscopic measurements ($z$\_sp). The colors on the five similar
plots are different $r$ magnitude bins. The top panels show
photo-$z$ estimates using only the optical $griz$ (left) and $ugriz$
(right) data. Adding the $u$-band data significantly improves the
estimates, but in both analyses large groups of outliers exist in
the $0.4 <$ $z$\_ML $< 0.6$ range. By adding the GALEX data (middle
panels), this group of outliers is removed, and the photo-$z$
predictions fall much closer to the spectroscopic measurements. Note
that BPZ consistently results in an excess of galaxies in the lowest
$z$ bins, so the scatter and bias are quoted in the legends both for
all galaxies and for those with photo-$z > 0.03$. In the lower left
panel, we show the SDSS ANNz photo-$z$ estimates for comparison. The
ANNz technique does result in less scatter than adding the GALEX
data; however, ANNz is being compared to a subset of its redshift
training and validation set, as opposed to the ML analysis, which
does not utilize any prior redshift information. The lower right
panel compares the redshift distributions from the five photo-$z$
analyses (colors) with the spectroscopic measurements (black).}
\label{fig:ML_comparisons}
\end{centering}
\end{figure*}

The addition of GALEX data to the optical measurements alleviates
the redshift-brightness degeneracy and greatly improves the
photo-$z$ estimation. In Fig.~\ref{fig:ML_comparisons}, the
upper-left panel shows the ML recovered photo-$z$ when using only
$griz$ data. As expected, the number of catastrophic failures is
high, resulting in a large standard deviation of $\sigma_z= 0.17
(1+z)$. Addition of the $u$ band data (upper-right panel) reduces
the number of catastrophic failures and halves the standard
deviation to $\sigma_z=0.08 (1+z)$.\footnote{We note that the
standard deviation of the $ugriz$ analysis is reduced by a 25\% to
$\sigma_z=0.06 (1+z)$ if SDSS {\em model} magnitudes are used. These
are the best internally calibrated magnitudes for SDSS photometry,
however, the model magnitudes cause the photo-$z$ bias to increase
by more than an order of magnitude when used with the GALEX data.}
Including the GALEX data (middle panels) reduces the standard
deviation to $\sigma_z=0.04 (1+z)$ and removes nearly all
catastrophic failures. Note that the addition of $u$ data has
negligible effect when the GALEX bands are added. The bottom-left
panel shows the comparison with ANNz. The standard deviation of ANNz
is $\sim$30\% smaller than GALEX + $griz$; although, we note that
ANNz is being compared to a sample that includes its own training
and validation set, which makes the comparison a bit unfair. We find
that simply adding the GALEX moderate exposures to $griz$ imaging
and using ML analysis techniques with 6 empirically motivated galaxy
templates provides photo-$z$ approaching the accuracy of ANNz on
$ugriz$.

\begin{figure}
\begin{centering}
\includegraphics[width=0.91\columnwidth]{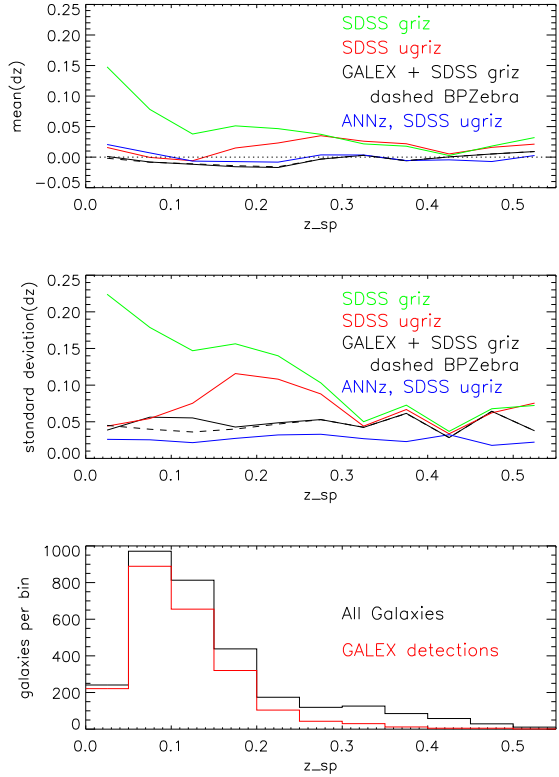}
\caption{Errors in photo-$z$ estimation versus spectroscopic
redshift. The mean (top) and standard deviation (middle) of $dz$ are
shown as a function of redshift. We compare the ML photo-$z$ results
using the SDSS $griz$ data (green), which is representative of the
BCS measurements, as well as the SDSS $ugriz$ data (red), the GALEX
+ $griz$ data (black), and the SDSS ANNz results (blue). The GALEX +
$griz$ data approaches the standard deviation of the ANNz results
without the use of priors or training sets. We also show the
improvement that can be achieved in the GALEX + $griz$ analysis by
using the ZEBRA code for $z_{ph} < 0.25$ and BPZ code otherwise,
which we call BPZebra (black dashed). In the bottom panel, the total
number of sources in each $z$ bin is shown (black) as well as the
total number of sources with a GALEX detection (red). }
\label{fig:zerr_z}
\end{centering}
\end{figure}

\begin{figure}
\begin{centering}
\includegraphics[width=0.91\columnwidth]{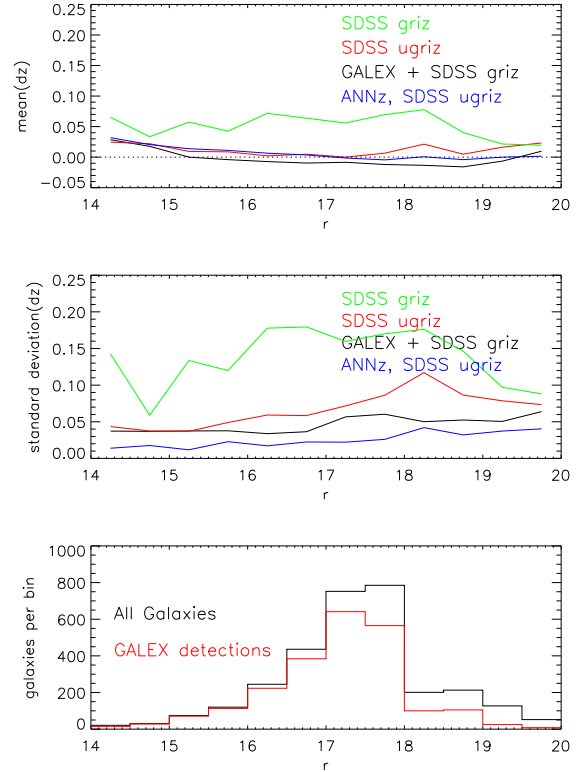}
\caption{Errors in photo-$z$ estimation versus source $r$ magnitude.
The mean (top) and standard deviation (middle) of $dz$ are shown in
different $r$ bins. (Colors are the same as Fig.~\ref{fig:zerr_z}
and \ref{fig:zerr_gr}.) The GALEX + $griz$ data approaches the
standard deviation of the ANNz results without the use of priors or
training sets. At the bottom, the total number of sources in each
$r$ bin is shown (black) as well as the total number of sources with
a GALEX detection (red). } \label{fig:zerr_r_mag}
\end{centering}
\end{figure}

\begin{figure}
\begin{centering}
\includegraphics[width=0.91\columnwidth]{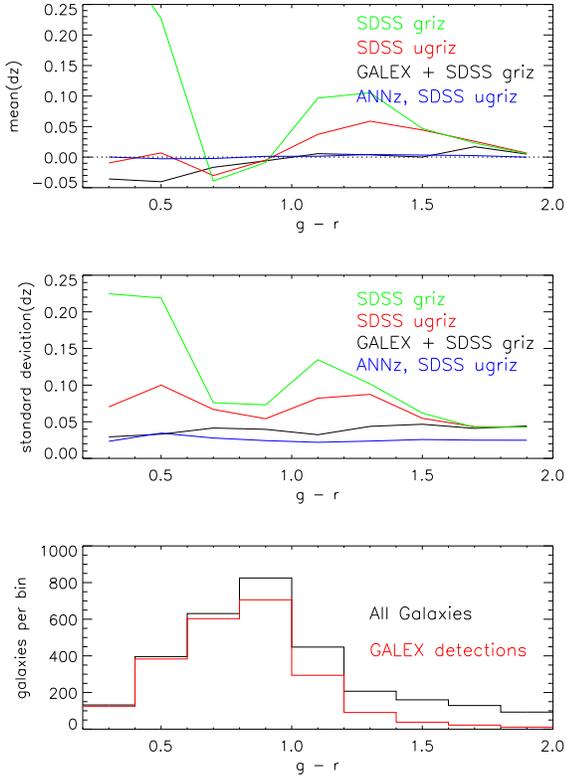}
\caption{Errors in photo-$z$ estimation versus color, $g-r$. The
mean (top) and standard deviation (middle) of $dz$ are shown in
different $g-r$ bins. (Colors are the same as Fig.~\ref{fig:zerr_z}
and \ref{fig:zerr_r_mag}.) For the low $g-r$ bins, or blue galaxies,
the standard deviation of the GALEX + $griz$ results are a huge
improvement over the SDSS only data and are essentially equivalent
to the ANNz results without the use of priors or training sets. At
the bottom, the total number of sources in each $g-r$ bin is shown
(black) as well as the total number of sources with a GALEX
detection (red). } \label{fig:zerr_gr}
\end{centering}
\end{figure}

We explore the performance of the photo-$z$ in more detail in
Fig.~\ref{fig:zerr_z}, \ref{fig:zerr_r_mag}, and \ref{fig:zerr_gr}
to investigate the dependence on redshift, magnitude, and color,
respectively. In Fig.~\ref{fig:zerr_z} we show how the photo-$z$
bias and scatter evolve as a function of redshift. Adding the GALEX
data dramatically reduces the bias and scatter over the optical
bands alone at $z<0.3$, beyond which the proportion of galaxies with
GALEX detections falls off at the current GALEX observation depths
(Fig.~\ref{fig:zerr_z}, bottom panel). Still, the performance
approaches the level of ANNz up to near $z \approx 0.4$. In
Fig.~\ref{fig:zerr_r_mag} we show the photo-$z$ bias and scatter as
a function of the source $r$ magnitude. Both remain nearly flat in
the regime $r < 19$, above which the fraction of galaxies detected
by GALEX falls to less than 1/2. These plots clearly indicate that
with deeper GALEX exposures, we can expect to improve our results
for fainter objects and higher redshifts. In Fig.~\ref{fig:zerr_gr}
the photo-$z$ performance as a function of galaxy color is examined.
The scatter is equivalent to (or possibly even lower than) ANNz for
$g-r < 0.6$ and is only slightly larger up to $g-r \approx 2$. When
compared to the other ML methods without GALEX photometry, the
addition of GALEX bands returns significantly more accurate
photo-$z$ for colors as red as $g-r = 1.4$.

We also consider removal of the excess of galaxies in the lowest BPZ
redshift bins ($z<0.03$). Cutting these galaxies results in a
$\sim$5\% reduction of the standard deviation and almost a factor of
ten reduction in bias (Fig~\ref{fig:ML_comparisons}, middle panels).
At $z_{ph}<0.25$ ZEBRA photo-$z$ scatter is $\sim$8\% smaller than
BPZ and does not show the excess in the lowest $z$ bin. A hybrid
technique (ZEBRA at $z_{ph}<0.25$ and BPZ at $z_{ph}>0.25$, which we
call BPZebra) can be used to take advantage of the fact that ZEBRA
does not have a pile-up of galaxies at low-$z$, which reduces the
total scatter by $\sim$8\% and reduces the total bias by a similar
amount to cutting the low-$z$ galaxies (dashed line in
Fig.~\ref{fig:zerr_z}).

\subsection{ANNz analysis for improving stripe 82 photo-$z$}
\label{sec:ANNzG_results}

The performance of ANNz using GALEX data (ANNzG, \S
\ref{sec:ANNz_anal}) is explored with several combinations of SDSS
bands.\footnote{We note that the training and validation sets are a
random sub-sample of 36\% of the catalog.} The best performance is
(not surprisingly) obtained with all five SDSS bands and GALEX
(ANNzG: $ugriz$); in this case the scatter is $\sigma_z = 0.018
(1+z)$. Removing the SDSS $u$ (ANNzG: $griz$) or $u$ and $z$ (ANNzG:
$gri$) bands only causes slight degradations in the photo-$z$
scatters to $\sigma_z \approx 0.020 (1+z)$. As a systematic test of
our ANNzG approach, we run the same analysis on the SDSS $ugriz$
data in our catalog and find that it gives $\sigma_z = 0.026 (1+z)$,
which is consistent with the scatter from the \citet{ANNz_SDSS_07}
ANNz pipeline on this data set of $\sigma_z \approx 0.027 (1+z)$.
These results combined with the results in \S \ref{sec:SDSS_spec}
indicate that the GALEX bands provide superior redshift information
to the SDSS $u$ and/or $z$ bands.

\begin{figure}
\begin{centering}
\includegraphics[width=0.9\columnwidth]{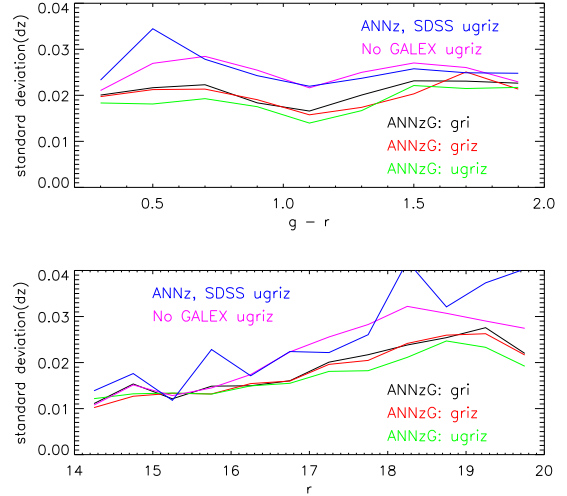}
\caption{Artificial neural network photo-$z$ scatter as a function
of galaxy color, $g-r$ (top panel), and magnitude, $r$ (bottom
panel), analyzed as described in \S \ref{sec:ANNz_anal} for
different combinations of SDSS and GALEX data. The scatter is
compared to the performance of the \citet{ANNz_SDSS_07} photo-$z$
pipeline (ANNz, SDSS $ugriz$, blue) on the same data set. The
addition of GALEX data to $ugriz$ data (ANNzG: $ugriz$, green)
provides the best photo-$z$ predictions, while GALEX combined with
$griz$ (ANNzG: $griz$, red) and even just $gri$ (ANNzG: $gri$,
black) only results in slight increases in scatter compared to the
complete data set. This indicates that the GALEX data provides more
redshift information than the SDSS $u$ or $z$ bands. As a systematic
check, we have run identical ANNz analysis on the SDSS $ugriz$ data
without GALEX (No GALEX $ugriz$, magenta), and we find that the
scatter distribution is similar to that recovered by
\citet{ANNz_SDSS_07}, both as a function of $g-r$ and $r$. The
galaxy distributions in the scatter bins are those shown in the
bottom panels of Fig.~\ref{fig:zerr_r_mag} and
\ref{fig:zerr_gr}.}\label{fig:Annz_gr_errors}
\end{centering}
\end{figure}

In Fig.~\ref{fig:Annz_gr_errors} we explore the color, $g-r$, and
magnitude, $r$, dependence of the photo-$z$ scatter from the ANNzG
and ANNz calculations. The addition of the GALEX data results in a
clear and significant reduction in scatter for nearly all color and
magnitude bins, with the exception of the reddest (high $g-r$) and
brightest (low $r$) galaxies. The general consistency between trends
in the ANNz pipeline results (ANNz, SDSS $ugriz$ in legend) and our
own analysis applied to the SDSS only data (No GALEX $ugriz$ in
legend) is an indication that our ANNzG results are robust. The
photo-$z$ bias was also explored, and it was found to be roughly
five to twenty times smaller than the scatter in each bin, so we do
not discuss it further.

\subsection{Public Photo-$z$ Catalogs} \label{sec:Photoz_catalog}

\begin{figure}
\begin{centering}
\includegraphics[width=0.9\columnwidth]{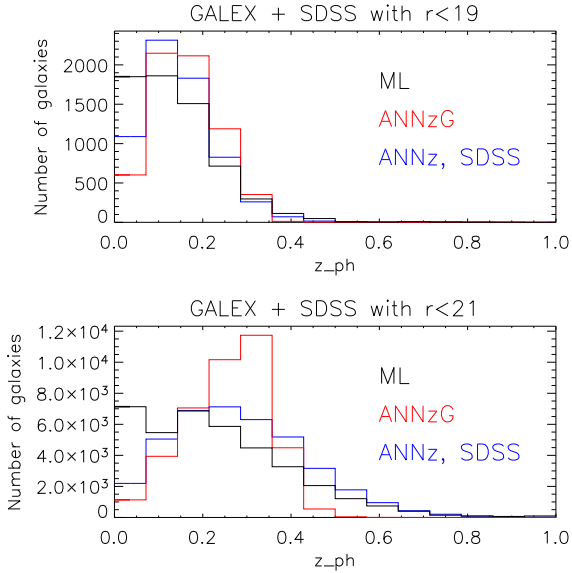}
\caption{Comparison of photo-$z$ distributions for SDSS data with
GALEX detections. The top panel shows data with $r<19$. The ML
analysis (black), the ANNzG analysis (red), and the SDSS ANNz
results (blue) are all relatively similar in this magnitude regime.
The bottom panel compares the same three analyses on data with
$r<21$. The ANNzG analysis clearly fails here, because the current
training set utilizes SDSS spectroscopic measurements, which have a
lower magnitude distribution (Fig.~\ref{fig:zerr_r_mag}, bottom
panel). The ML analysis distribution has some failures as well, but
it has significantly less bias than the ANNzG analysis. The excess
in the lowest $z$ bin has been observed in all ML analyses with
GALEX data (Fig.~\ref{fig:ML_comparisons}). In our public catalogs,
the ANNzG and ML results are provided for all galaxies with $r<19$,
while ML results are provided for all galaxies detected by GALEX
with $r<21$.} \label{fig:SDSS_ANN_BPZ_compare}
\end{centering}
\end{figure}

Here we apply our ML (\S \ref{sec:SED_photoz_anal}) and ANNzG (\S
\ref{sec:ANNz_anal}) approaches for calculating photo-$z$ to stripe
82 galaxies that do not have spectroscopic data. The redshift
distributions from these analyses as well as the SDSS ANNz pipeline
are compared in Fig.~\ref{fig:SDSS_ANN_BPZ_compare}. The top and
bottom panels show the redshift distributions for galaxies with
$r<19$ and $r<21$, respectively. Because of the lack of SDSS
spectroscopic observations for GALEX detected galaxies with $r>19$
(Fig.~\ref{fig:zerr_r_mag}) and $z>0.3$ (Fig.~\ref{fig:zerr_z}), the
current ANNzG analysis does not have accurate training above this
limit. The excess number of galaxies at $z \approx 0.3$ in the lower
panel of Fig.~\ref{fig:SDSS_ANN_BPZ_compare} is due to the resulting
failure of the ANNzG analysis for galaxies with $r>19$. As expected,
this clearly indicates that to use empirical photo-$z$ techniques
one must ensure that spectroscopic training sets are representative
of the complete photometric sample.

The ML analysis, on the other hand, does not require spectroscopic
training, and the lower panel of Fig.~\ref{fig:SDSS_ANN_BPZ_compare}
shows that the ML analysis on galaxies detected by GALEX produces a
similar distribution to the SDSS ANNz pipeline even for dimmer
galaxies with $19<r<21$. The primary difference between the ML and
ANNz distributions is the excess in the ML lowest redshift bin,
which is a known failure of the BPZ code used for this analysis
(described in \S \ref{sec:SDSS_spec}).

The resulting catalogs containing both the ML and ANNzG analyses for
galaxies observed by GALEX will be made publicly available for the
community at www.ice.csic.es/personal/jimenez/PHOTOZ. The catalogs
will be updated with more complete versions as our GALEX stripe 82
observations are completed. Since the ANNzG analysis has only been
trained up to $r\approx19$, we provide an ANNzG photo-$z$ catalog
for all SDSS galaxies up to this limit as well as ML photo-$z$ on
those same galaxies. Since the ML analysis does not require
spectroscopic training, we also provide a catalog with ML photo-$z$
estimates for all SDSS galaxies that have GALEX detections and
$r<21$.

\section{Conclusions}

In order to obtain accurate photometric redshifts as efficiently as
possible for the areas surveyed by SZ experiments, we have obtained
moderate-depth GALEX photometry. With a modest observing campaign,
and using already available MIS observations, we have already
covered an area of $\sim$60 deg$^2$ to a mean depth of $\sim$3 ks.
At the completion of our $\sim$210 ks of observations, we will have
covered $\sim$100 deg$^2$ to this depth.

\cite{Budavari_photoz_2005} previously used $ugriz$ SDSS DR1
photometry together with GALEX Medium Imaging Survey (MIS, 1.4 ks
exposure) $F_{UV}$ and $N_{UV}$ photometry to determine photo-$z$
for about 10000 galaxies up to $z \approx 0.25$. They use an
empirical technique which relies on a training set of about 6000
objects, and obtained photo-z errors of $\sigma_z=0.026$ on the
training set, which is similar to the SDSS ANNz performance. As
large training sets and $u$-band data may not be available for the
next generation large-area SZ cluster surveys and as $u$-band
photometry may not be available for future optical surveys such as
BCS, DES, and LSST, we have considered two cases.

To be independent of training sets, we considered a
spectral-energy-distribution, or template-based, photo-$z$ approach.
As we have found that suitable templates for use with GALEX
observations were not publicly available, we have constructed new,
physically motivated, spectral templates. They are publicly
available at www.ice.csic.es/personal/jimenez/PHOTOZ.

Using the SDSS spectroscopic survey we have shown that the addition
of GALEX photometry to only $griz$ bands makes possible the use of
simple maximum likelihood techniques, without resorting to  Bayesian
priors. This approach obtains $\sigma_z = 0.04(1+z)$ for $r<20$
galaxies, which includes luminous galaxies up to $z \approx 0.4$.
This accuracy approaches that obtained using spectroscopic training
of neural networks on $ugriz$ photometry of the same galaxy sample.
In particular, we have shown that the large number of catastrophic
failures that occur for $griz$-based and $ugriz$-based maximum
likelihood photo-$z$ determinations is nearly eliminated by adding
UV photometry from GALEX data to $griz$ data. The improvement is
especially notable for blue galaxies; for galaxies with $g-r < 0.6$,
we obtain photo-$z$ scatter of $\sim$$0.03 (1+z)$. We find that the
addition of UV observations to $griz$ photometry, provides
significantly better photo-$z$ than the addition of $u$-band
observations.

Beyond $z \approx 0.4$, the GALEX $\sim$3 ks exposures do not have a
sufficient number of detections to dramatically improve the $griz$
observations. Clearly, moderately deeper observations would help to
bring the utility of GALEX observations closer to $z \approx 1$. We
note that the current depth of $z \approx 0.4$ looks back through
roughly 33\% of the age of the universe and samples a volume of 15
Gpc$^3$. Maybe more importantly, $\sim$20\% of the clusters that
will be detected by the SZ experiments (above a dark matter mass of
$3 \times 10^{14}$ M$_{\odot}$) are at $z < 0.4$. If redshift up to
$z=1$ were accessible by GALEX, $\sim$60\% of the age of the
universe and a volume of 153 Gpc$^{3}$ would be surveyed; 86\% of
the clusters that will be detected by the SZ experiments (above a
dark matter mass of $3 \times 10^{14}$ M$_{\odot}$) and 90\% of the
resolved ones, are expected to be at $z < 1$.

The most important aspect of the results presented here, is that the
photo-$z$ accuracy of $\sigma_z =0.04(1+z)$ at $z< 0.4$ was obtained
using only maximum-likelihood fits to six galaxy templates in BPZ,
without resorting to priors or training-sets. As the acquisition of
training sets or priors relies on obtaining large spectroscopic
data-sets, we consider the moderate GALEX exposures an efficient way
to obtain accurate photo-$z$ over large areas.\footnote{Note that we
just integrated $\sim$2.4 days and that, for example, a program 10
times longer could provide photo-$z$ for about 1000 deg$^2$, which
(we estimate) is the optimal area to extract cosmological
information from SZ surveys} Further, GALEX photometry gives a
direct measurement of star formation and AGN activity
\citep{AtleeGould07}, a subject that we are continuing to explore.

Should large spectroscopic training sets be available, we have
considered the effect of adding UV photometry to optical data on the
performance of an artificial neural network photo-$z$ calculation.
The addition of GALEX observations to optical $griz$ (or even just
$gri$) observations yields photo-$z$ that have $\sigma_z = 0.02
(1+z)$, which is $\sim$30\% smaller scatter than was obtained on the
same data set using only SDSS $ugriz$ observations.

We make our photo-$z$ catalogs of stripe 82 galaxies detected by
GALEX publicly available at www.ice.csic.es/personal/jimenez/PHOTOZ.
The catalogs contain the results of the ML photo-$z$ calculation for
all GALEX detected galaxies with $r<21$ as well as the ANNzG and ML
photo-$z$ calculations for all SDSS stripe 82 galaxies in GALEX
fields with $r<19$. The posted catalogs will be updated as our GALEX
observations and analysis are completed.

The approach proposed here can provide a useful catalog for
weak-lensing studies as photo-$z$ remain accurate for the bluest
galaxies. These determinations are commonly the most difficult to
obtain because spectra of blue galaxies in the optical bands show an
almost featureless power law spectral energy distribution. We
envision our SED-based ML approach to be useful for
cross-correlation studies with CMB maps, where deep photo-$z$ are
needed over large areas  and large training sets may not be
available. Possible applications of these studies include improving
our understanding of dark energy using cluster counting techniques
\citep{SPT_white,lima-hu07}, the kSZ effect \citep{carlosksz}, and
the lensing of the CMB by large-scale structure \citep{melita}.

\section*{Acknowledgements}

MN is grateful to Robert Lupton, John Hughes, Yen-Ting Lin, and
Chris Hirata for their useful and informative discussions and
suggestions as well as Peter Friedman for help with the GALEX data
and his advisor, Suzanne Staggs, for supporting his work on this
project. RJ and LV thank Txitxo Ben{\'{\i}}tez for helpful
discussions. The research of MN is supported by the U.S. National
Science Foundation through grant PHY-0355328 as well as a Princeton
University Centennial Fellowship. RJ acknowledges support from
FP7-PEOPLE-2007-4-3-IRG grant and CSIC I3 grant 200750I037. LV
acknowledges support of FP7-PEOPLE-2007-4-3-IRGn202182  and CSIC I3
grant 200750I034. RJ and LV are partially supported by GALEX grant
GI3-095. RJ and DNS are partially supported by NSF grant
PIRE-0507768.

Funding for the creation and distribution of the SDSS Archive has
been provided by the Alfred P. Sloan Foundation, the Participating
Institutions, the National Aeronautics and Space Administration, the
National Science Foundation, the US Department of Energy, the
Japanese Monbukagakusho, and the Max Planck Society. The SDSS Web
site is www.sdss.org. The SDSS is managed by the Astrophysical
Research Consortium (ARC) for the Participating Institutions. The
Participating Institutions are the University of Chicago, Fermilab,
the Institute for Advanced Study, the Japan Participation Group, The
Johns Hopkins University, the Korean Scientist Group, Los Alamos
National Laboratory, the Max Planck Institute for Astronomy (MPIA),
the Max Planck Institute for Astrophysics (MPA), New Mexico State
University, the University of Pittsburgh, Princeton University, the
United States Naval Observatory, and the University of Washington.

The Galaxy Evolution Explorer (GALEX) is a NASA Small Explorer
(www.galex.caltech.edu). The mission was developed in cooperation
with the Centre National d'Etudes Spatiales of France and the Korean
Ministry of Science and Technology.

\end{document}